\documentclass[epj,twocolumn,galleyproof,superscriptaddress]{revtex4-1}
\usepackage{amsmath,amssymb}
\usepackage{graphicx,color,dsfont}
\usepackage{etoolbox}

\begin{document}

\title{From seconds to months: multi-scale dynamics of mobile telephone calls }
\date{\today}

\author{Jari Saram\"aki}\affiliation{Complex Networks Group, Aalto University School of Science, P.O. Box 12200, FI-00076, Finland}
\author{Esteban Moro}\affiliation{Departamento de Matem\'aticas \& GISC, Universidad Carlos III de Madrid, Legan\'es 28911, Spain}

\begin{abstract}
Big Data on electronic records of social interactions allow approaching human behaviour and sociality from a quantitative point of view with unforeseen statistical power. Mobile telephone Call Detail Records (CDRs), automatically collected by telecom operators for billing purposes, have proven especially fruitful for understanding one-to-one communication patterns as well as the dynamics of social networks that are reflected in such patterns. We present an overview of empirical results on the multi-scale dynamics of social dynamics and networks inferred from mobile telephone calls. We begin with the shortest timescales and fastest dynamics, such as burstiness of call sequences between individuals, and "zoom out" towards longer temporal and larger structural scales, from temporal motifs formed by correlated calls between multiple individuals to long-term dynamics of social groups. We conclude this overview with a future outlook.
\end{abstract}
\maketitle


\section{Introduction} 

Electronic records have revolutionised studies of human social behaviour. Instead of having to rely on field observations or costly questionnaire-based surveys, today's social scientists can follow social interactions between millions of individuals with the help of email and social media logs as well as CDRs (Call Detail Records) extracted from billing systems of mobile telephone operators. In addition to social scientists, this data-driven movement has attracted large numbers of physicists, interested in a variety of topics such as collective behaviour and emergent network structures. The term \emph{computational social science} has been coined to describe this new field of inquiry~\cite{Lazer2009}.

Data sets on mobile telephone calls have certain advantages over other sources for studying social behaviour. First, mobile telephones are ubiquitous and used by all age groups and in all social strata, whereas the user base of, say, Twitter cannot yet be considered as representative of the general population. Second, a phone call needs to be picked up before its details are recorded as CDRs by the operator  (caller, callee, time, duration). Hence, CDRs are records of verified, time-stamped one-to-one communication. This greatly facilitates constructing social networks from the data, and especially allows for temporal analysis of communication patterns, to the contrary of \emph{e.g.}~emails where recipient lists may be long and where there is no guarantee when (or if!) an email has been actually read. 

Because of the above, mobile phone call records have been used in numerous studies on diverse topics~\cite{Krings2015}: 
social network structure (e.g.~\cite{Onnela2007,Eagle2009}), geography of social relationships (e.g.~\cite{Lambiotte2008,krings2009}), disaster response (e.g.~\cite{Lu2012}), economic development (e.g.~\cite{Eagle2010}), and human mobility patterns (e.g.~\cite{gonzalez2008}), to name a few. In the earliest investigations, it was typical to aggregate calls between individuals over time and treat the resulting networks as static~\cite{Onnela2007}, or to consider slow dynamical processes such as dynamics of social groups being formed and merged~\cite{Palla2007}. Recently, however, there has been increased interest in dynamics on multiple time scales -- time stamps of individual calls, statistics of inter-call times, and their network-level consequences have become focal topics (e.g.~\cite{Candia2008,Karsai2011,Miritello2011}). This is both because of the rich dynamics observed in empirical data, and because of the added level of detail for understanding human behavioural patterns. At the same time, there has been a general increase of interest in \emph{temporal networks}, networks that consist of nodes that are connected by \emph{events} or \emph{contacts} only at specified times~\cite{Holme2012}.

\begin{figure*}
\begin{center}
\includegraphics[width=0.75\linewidth]{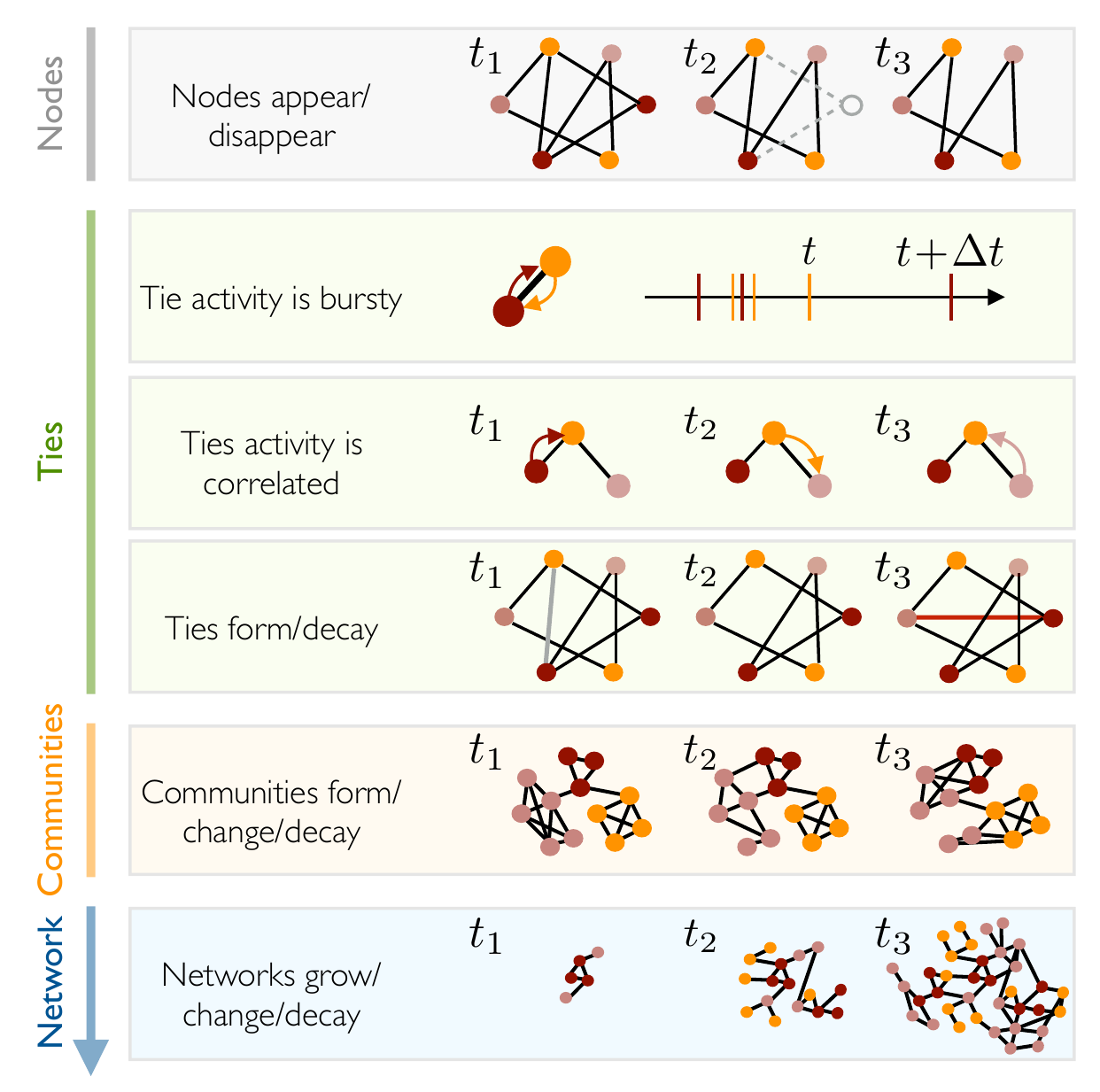}
\caption{An overview of temporal and structural scales in mobile call networks, from activity on short time scales at the level of ties and nodes to slower dynamics at mesoscopic and network-wide scales.}\label{fig:timescales}
\end{center}
\end{figure*}

In this paper, we attempt to provide a brief overview of what is known to go on in mobile telephone communication and related social networks at different time scales, from short to long. This range of time scales is also related to structural scales, as illustrated in Figure~\ref{fig:timescales}. At the shortest time scales the focus is on the timings of individual calls and their correlations, and the relevant structural units are nodes and ties.
Moving on to dynamics on longer time scales, the focus gradually shifts to sets of ties, such as egocentric networks -- sets of ties surrounding an individual -- and social groups and communities. Finally, there is dynamics at the level of entire networks. A future outlook -- where the field is heading, and where should it be heading -- is given at the end.

\section{How to construct social networks from CDRs}

Before we address the network dynamics that are at the focus of this article, let us briefly explain how social networks are constructed from CDRs.

Typically, the source data consists of entries containing at least the following items: caller id, callee id, time of event, duration of event (if any), event type (e.g.~call, text message or multimedia message). These entries span some range of time, \emph{e.g.} a month or six months. Often, the data has been filtered to only contain ids of customers of the source operator, and perhaps only those of private subscribers (non-company users). The ids are typically hashed versions of the original phone numbers -- surrogate keys -- generated for privacy reasons at or close to the data source.

An unweighted, static social network can then be constructed simply by considering the ids as nodes, and connecting two nodes by links if there are call or message events between them in the data. Here, some filtering is usually applied, \emph{e.g.} by requiring that there are one or several calls from $i$ to $j$ and vice versa for an $i-j$ link to exist. One may also consider link \emph{weights}, that is, social tie strengths, computed either as total numbers of calls or total call duration between two individuals~\cite{Onnela2007}. Obviously, for such static aggregated networks, the time span covered by the data has an effect on the outcome~\cite{krings2012effects}.
The simplest way of constructing dynamic networks is then to split the data into consecutive time windows and apply the above procedure to each window, yielding a discrete time series of time-dependent links that may be weighted. For the most fine-grained dynamics, the concept of links is practically discarded as links are only considered a substrate for communication events. The events themselves form temporal networks, where callers and recipients are linked by a time-stamped contact only at those time points when there is a call (or message) in the CDR data. This is the case when activity patterns are studied at nodal or tie level.

\section{Activity patterns at the level of nodes and ties}

\subsection{Human communication is bursty}

\begin{figure}
\begin{center}
\includegraphics[width=0.95\linewidth]{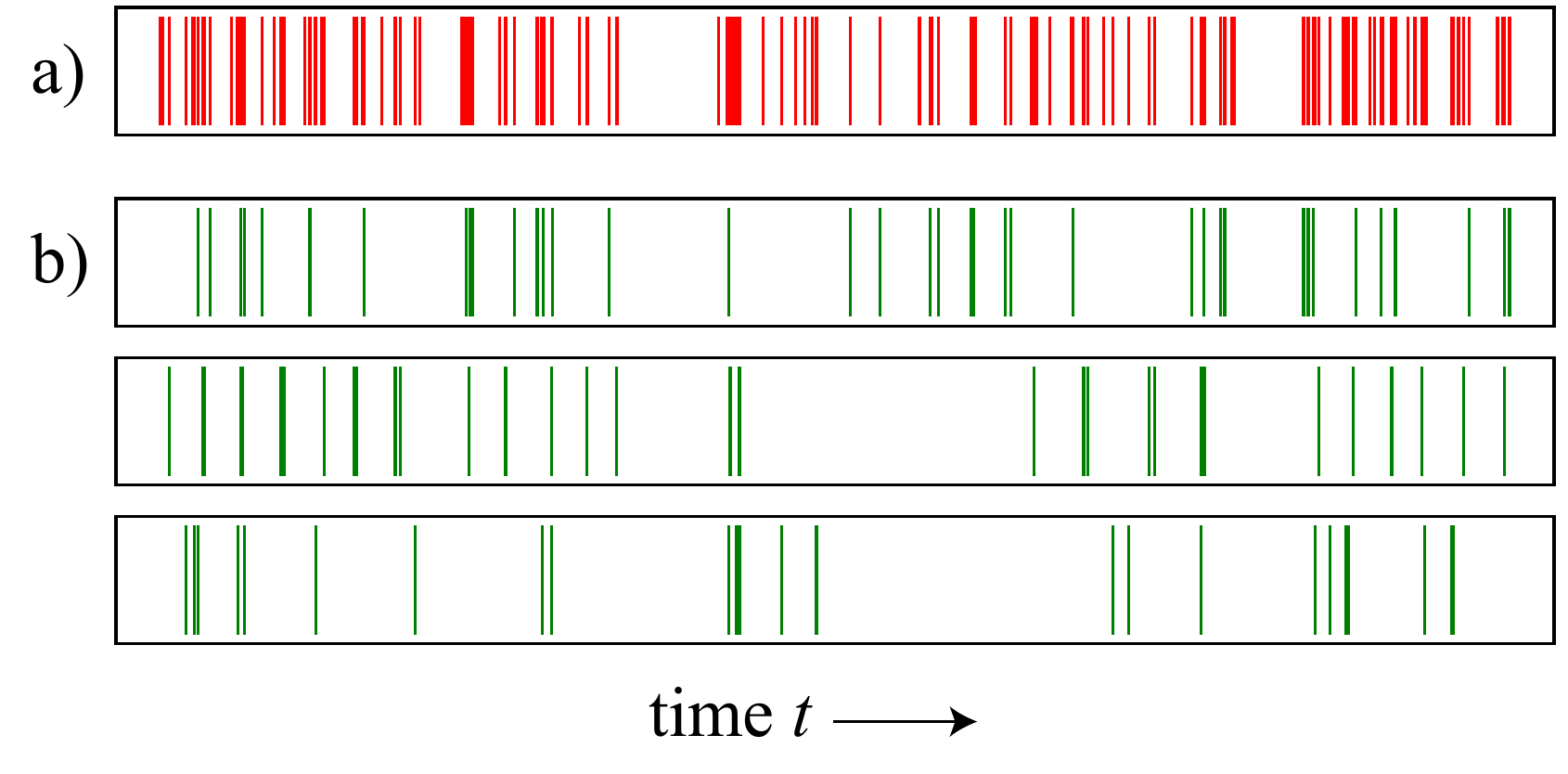}
\caption{a) Timeline of all outgoing calls of one individual for one month, b) timeline of calls from the same individual to his/her three top acquaintances. In network terms, a) represents the timeline of a node, and b) the timelines of links (social ties).  Data from~\cite{Saramaki2014}, figure after~\cite{Karsai2012b})}\label{burstiness_schematic}
\end{center}
\end{figure}

Let us now begin our journey from the smallest towards the largest by looking at the very atoms of communication, the individual communication events. Since we are interested in the time domain, it is then natural to focus on the timings of events: what can we say about their properties and statistics? It has become apparent in the recent years that human activity in general is rarely uniformly randomly distributed in time. Instead, human activity patterns are commonly \emph{bursty}~\cite{Barabasi2005} -- there are rapid bursts of successive events that are separated by longer periods of inactivity. Mobile telephone calls are not different: time series of calls are typically bursty, and accordingly,  the distribution of times between calls is heavy-tailed~\cite{Vazquez2007,Candia2008,Iribarren2009,Karsai2011}. Fig.~\ref{burstiness_schematic} displays an example of this alternation between bursts of calls and periods of no communication, for one individual and three of his/her social ties. All displayed time series are clearly bursty, with a high level of variance in times between successive calls.

\begin{figure}
\begin{center}
\includegraphics[width=0.72\linewidth]{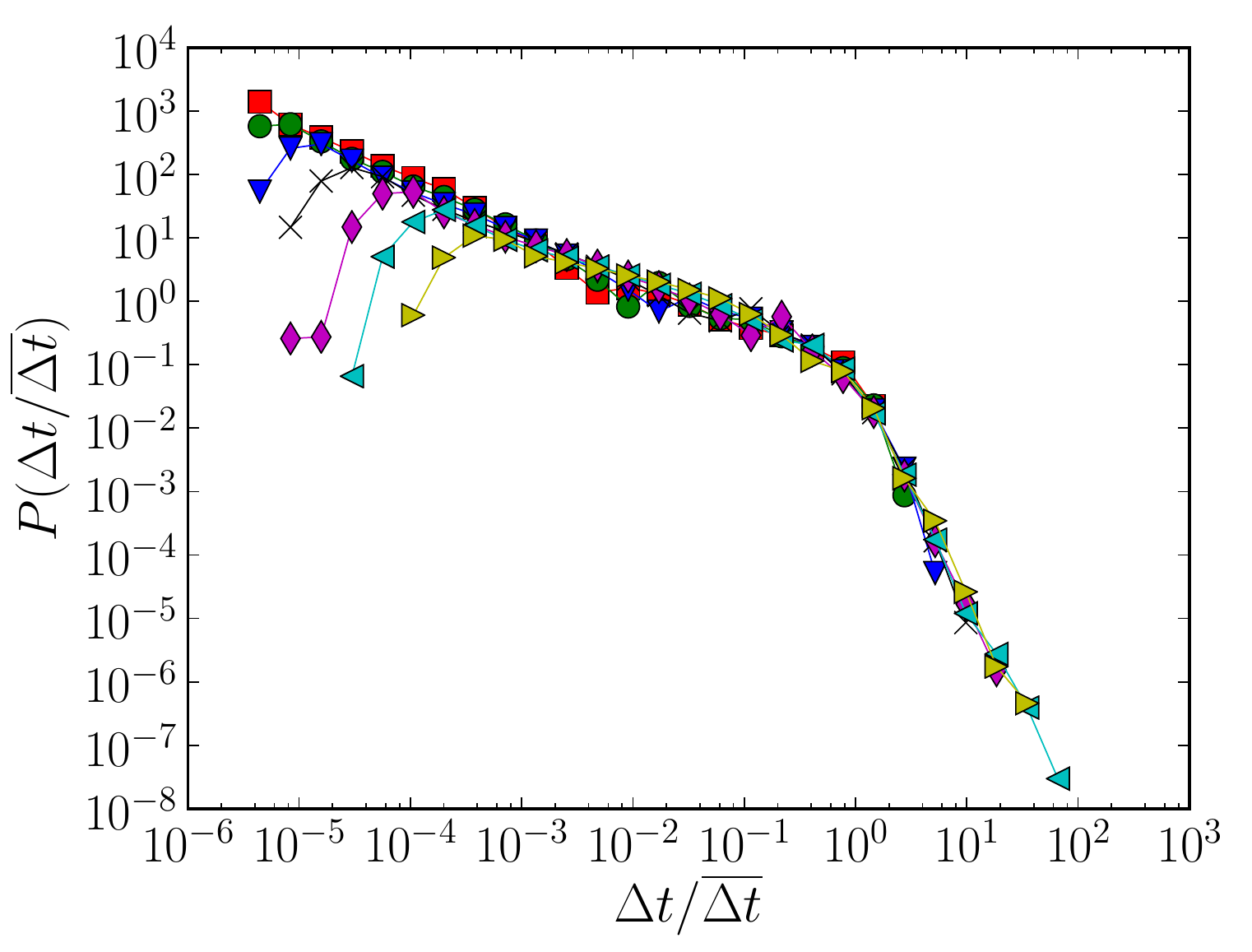}
\caption{Probability density functions of scaled inter-call times on edges, \emph{i.e.} between pairs of individuals. Because of the broad distribution of call activity levels of links, links have been grouped according to their number of calls. Then, for each group, the PDF of inter-event times $\Delta t/\overline{\Delta t}$ has been calculated separately, where $\overline{\Delta t}$ is the average inter-call time for the group.}\label{ietd}
\end{center}
\end{figure}


The level of burstiness in a time series can be measured with a single quantity 
$B=\left(\sigma_{\Delta t}-\overline{\Delta t}\right)/\left(\sigma_{\Delta t}+\overline{\Delta t}\right)$, where $\sigma_{\Delta t}$ is the variance and $\overline{\Delta t}$ the mean of inter-event times $\Delta t$~\cite{Goh2008}. However, it pays off to inspect the statistics of inter-call times in more detail.  At first, it would seem natural to look for burstiness in the statistics of call timings by simply inspecting the probability density function (PDF), $P(\Delta t)$, of all times $\Delta t$ between successive calls (either of an individual, or associated with one social tie). However, the result would be difficult to interpret, as it would arise from a mixture of inhomogeneities: it is known that the general activity levels of individuals and the number of calls on each of their links are also broadly distributed (see, \emph{e.g.}, \cite{Onnela2007,Onnela2007b,Saramaki2014}). Because of this, the typical way of characterising the statistics of inter-call times, introduced in~\cite{Candia2008}, is to first group either the individuals or their ties according to their total number of calls. Then, one can compute separately for each group a scaled version of the inter-call time PDF, $P(\Delta t / \overline{\Delta t})$, where $\overline{\Delta t}$ is the average inter-call time computed for the group in question. It has been observed that this results in data collapse, where the scaled PDFs for different groups  closely match, both for inter-call times of individuals~\cite{Candia2008} and inter-call times of ties, \emph{i.e.}~between pairs of individuals~\cite{Karsai2011,Miritello2011}.

Figure~\ref{ietd} displays the scaled inter-call time distributions for ties, computed for a subset of the data used in~\cite{Karsai2011}. There are three regions of interest. First, for low $\Delta t / \overline{\Delta t}$, there is no data collapse, indicating the existence of a time scale that does not depend on average inter-call times. This time scale has to do with repeated and forwarded calls, and we will return to this in Sec.~\ref{correlated}. The non-scaling region is followed by a power-law decay of inter-call times, indicating the presence of burstiness in the data. Finally, the PDF drops steeply; this can be associated with the effect of a finite observation period. See \cite{Kivela2014} for a discussion on estimating true inter-event time distributions from finite observation periods.

As seen above, the timings of calls are bursty both for individuals (nodes) and their social ties (links). Which elements, then, are the drivers of burstiness? Is link burstiness "inherited" from the burstiness of nodes, or is node-level burstiness merely a consequence of the links being bursty? In 
~\cite{Karsai2012b}, Karsai \emph{et al.} argue that the latter explanation is correct. Their argument is based on correlations within call sequences on links, that is, between pairs of individuals. The existence of such correlations at the nodal level was shown in~\cite{Karsai2012a} by considering the distribution of the numbers of events $E$ in \emph{bursty periods}, that is, trains of calls where each successive call takes place within some $\delta t$ time units. The distribution of event numbers $P(E)$ was seen to follow a power law in the original data, $P(E)\propto E^{-\beta}$, whereas for randomised reference data with shuffled inter-call times on links $P_{ref}(E)\propto e^{-E}$. This shuffled reference corresponds to a case where the inter-call time distribution of links is the same as in the original data, but all correlations have been removed. Since $P(E)\neq P_{ref}(E)$, there are correlations within the call trains. It should be noted that there are other suggested explanations; it has been argued based on e-mail data that burstiness results from the interplay of Poissonian processes and circadian and weekly patterns~\cite{Malmgren2008}.  Ref.~\cite{Jo2012b} claims that this is not the case, based on a procedure that removes the effects of such patterns from CDR data.

The underlying network structure can have dramatic effects on dynamics taking place on networks, and this is also true for temporal networks and their inhomogeneities. Especially, the effect of burstiness on spreading processes on temporal networks has been a hot research topic lately. The studied spreading processes include simple, deterministic Susceptible-Infectious (SI) spreading where contact events always transmit an "infection" from infectious to susceptible individuals, as well as more complex processes, such as threshold dynamics. Simulations of such processes on top of empirical contact sequences have shown that non-Poissonian inter-event times have effects on spreading dynamics in email networks~\cite{Vazquez2007,Iribarren2009}, call networks~\cite{Miritello2011,Karsai2011,Kivela2012,Backlund2014}, contact networks~\cite{Rocha2011}, and in various temporal network models~\cite{Rocha2013,Horvath2014}. For temporal networks of mobile telephone calls, the current understanding is that burstiness slows down network-wide spreading as compared to a reference case of Poissonian inter-call times~\cite{Karsai2011,Kivela2012}. However, it may also speed up the very early stages of spreading dynamics. The slowing-down because of burstiness has a simple explanation: high variance of inter-call times increases the expected waiting times on links. This is the classical waiting time paradox. Another way of viewing the effect of burstiness is to consider the latencies of temporal paths consisting of time-respecting sequences of calls~\cite{Pan2011}: temporal paths take longer to traverse when call sequences are bursty, and deterministic SI spreading by definition follows the fastest temporal paths.  However, the general picture of the effects of burstiness is still far from complete; there are conflicting results and special cases~\cite{Rocha2011,Horvath2014}. 



\subsection{There are temporally correlated call patterns}\label{correlated}

Moving beyond individual nodes and links towards larger network neighbourhoods, it is natural to expect that the timings of calls should reflect social behaviour in groups. Here, at the smallest level, one would expect to see timing correlations between calls to and from one individual's acquaintances, \emph{i.e.}~between calls on adjacent links. This is indeed the case -- such correlations are the reason for the non-scaling region in the distribution of inter-call times (Fig.~\ref{ietd},~\cite{Karsai2011,Miritello2011}). The lack of data collapse indicates a time scale measured in real units of time instead of group averages. The time scale in question is $\sim$20-30 seconds and it corresponds to the typical time it takes to return or forward a call (get a call and then call someone else). The peak around 20-30 seconds is clearly visible in the shapes of triggered time correlation functions  in Ref.~\cite{Backlund2014} ("density of preceding events"). Correlated timings of calls around individuals play a major role in the dynamics of threshold processes simulated on mobile call networks~\cite{Backlund2014}, and a lesser role in the dynamics of SI spreading~\cite{Kivela2012}. This has been seen with the help of temporal reference models where link-link correlations have been removed. For SIR (Susceptible-Infectious-Recovered) spreading with small to moderate transmission probability, such correlations facilitate spreading and cascades are larger than for a Poissonian reference~\cite{Miritello2011}.

It is natural to assume that often, there is some (causal) connection between subsequent calls involving the same individual, given that the calls follow one another within some short time difference $\Delta t$. Such a connection may be related to information transmission and forwarding, or received information triggering further calls ("you should call Mum."). Then, one may group subsequent calls into sets with the hope of finding patterns of interest. In Ref.~\cite{Kovanen2011}, Kovanen \emph{et al.} introduced the concept of \emph{temporal subgraphs} as a way of achieving this grouping. The concept of temporal subgraphs builds on the notions of $\Delta t$-adjacency and $\Delta t$-connectivity. Two calls are $\Delta t$-adjacent if they share a node and take place within $\Delta t$ time units, and two calls are $\Delta t$-connected if one can trace a path of $\Delta t$ adjacency between them. Temporal subgraphs are then defined as sets of $\Delta t$-connected calls.

Analogously to the concept of \emph{motifs}~\cite{Shen-Orr2002,Milo2002} in static networks, one may extract all different temporal subgraphs from a call network (given a choice of $\Delta t$) and group them into equivalence classes, \emph{temporal motifs}~\cite{Kovanen2011,Kovanen2012,Kovanen2013}. Then, the numbers of subgraphs in different classes provide information on temporal processes in the network -- e.g., ubiquitous chains and stars may reflect transmission and spreading of information. For temporal motifs, the equivalence classes are best defined on the basis of the order of calls: a temporal subgraph where A calls B who calls C should be equivalent to a subgraph where D calls E who calls F at some different point in time. Then, class equivalence can be addressed by mapping the subgraphs to directed, coloured graphs and then applying graph isomorphism techniques~\cite{Kovanen2011}.

\begin{figure}[!t]
\begin{center}
\includegraphics[width=0.95\linewidth]{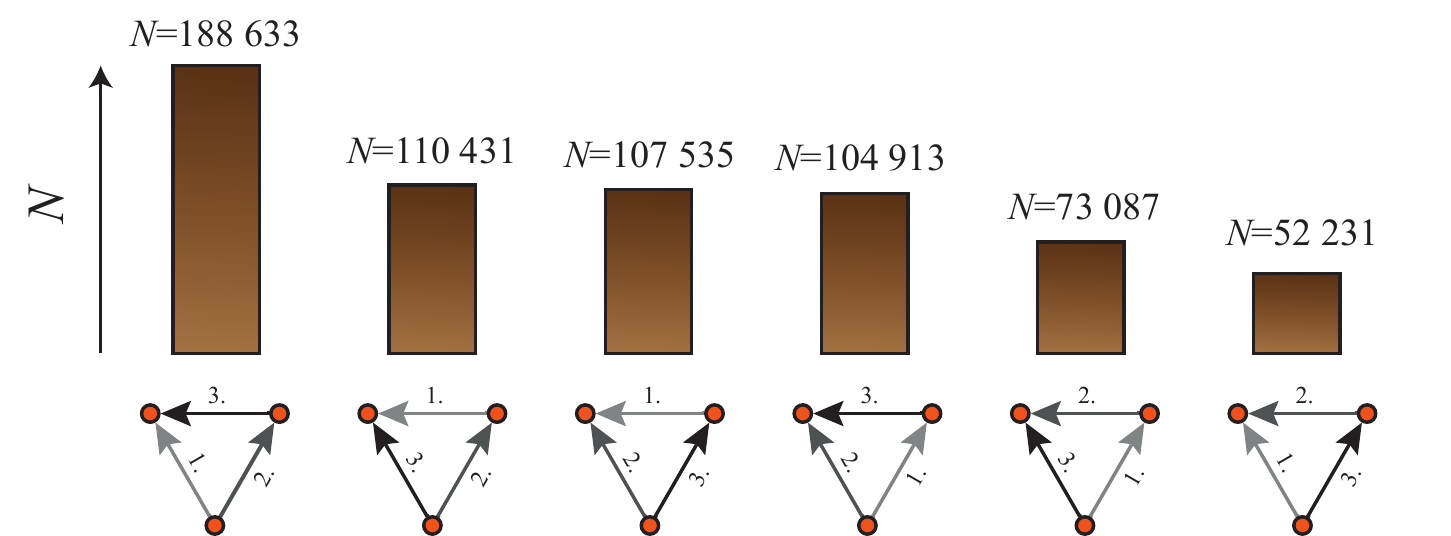}
\caption{Counts of triangular 3-call motifs in time-stamped mobile phone call data. After~\cite{Kovanen2012}.}\label{fig:motifs}
\end{center}
\end{figure}

For mobile call networks, temporal motif analysis of all 3-event motifs reveals that the most common ones reflect burstiness (three calls from A to B, A calling B who shortly thereafter calls A back , etc)~\cite{Kovanen2011}. When counts of triangular motifs are analysed (see Fig.~\ref{fig:motifs}), it is seen that the most common ones can have a direct causal explanation (e.g. A calls B and C, and then C calls B), whereas the least common ones appear to have arisen by random chance (A calls B, C calls B, A calls C). More detailed results are obtained when properties of the callers (gender, age) or links (intra- or inter-community) are taken into account: nodes in common temporal motifs tend to have similar properties (\emph{temporal homophily}), female motifs are different from male motifs (chains and stars vs.~"ping-pong"), and motifs within communities are more complex than those between~\cite{Kovanen2013}. 

\subsection{There are daily and weekly rhythms}\label{circadian}

In addition to the micro-level correlations between timings of calls discussed above, there are rhythms whose origins are in the behavioural patterns of individuals but  that are also clearly apparent at the aggregate level. In general, human activity follows a circadian rhythm, phase-locked to the day-night cycle, and this is also evident in call activity (see, e.g.,~\cite{krings2012effects, Jo2012}). Here, an interesting application is to consider the geospatial aspects of circadian rhythms, and measure call frequencies at different times of the day at different (tower) locations, which helps to understand spatiotemporal hotspots and the "rhythms of cities"~\cite{Louail2014}. Besides daily rhythms, there are also differences between weekdays -- not only in terms of call activity levels, but also in terms of who is being called~\cite{krings2012effects}: weekends are different from weekdays.


\section{Dynamics of ties}

\subsection{Mechanisms for tie creation and destruction}

Communication events are constituents of human relationships. At longer time scales (typically months), some social relationships are formed while others decay in time. The dynamics of ties is not random; several factors moderate their dynamics. First, there are meaningful social mechanisms behind link dynamics that have to do both with intention of individuals and stochastic elements. Sociological studies have revealed that many social mechanisms such as triadic closure (embeddedness), homophily, reciprocity, geographical proximity, or preferential attachment trigger the process of link formation (and conversely link decay) \cite{rivera2010dynamics}. For example, it is very likely that a link is formed between two persons who already share a wealth of common friends, or who happen to live near one another. In fact, these well-known mechanisms are behind most of the friend recommendation algorithms in electronic social networks \cite{liben2007link}. Second, the amount of social interactions that humans can handle is constrained: time, socio-economical status, and/or cognitive capacity do limit the number of social connections we can maintain in time \cite{dunbar1992neocortex,gonccalves2011modeling,miritello2013time}. This impacts the way how humans distribute communication between their connections, and also how humans balance the process of creating new links with that of destroying old ones.

\begin{figure}
\begin{center}
\includegraphics[width=0.99\linewidth]{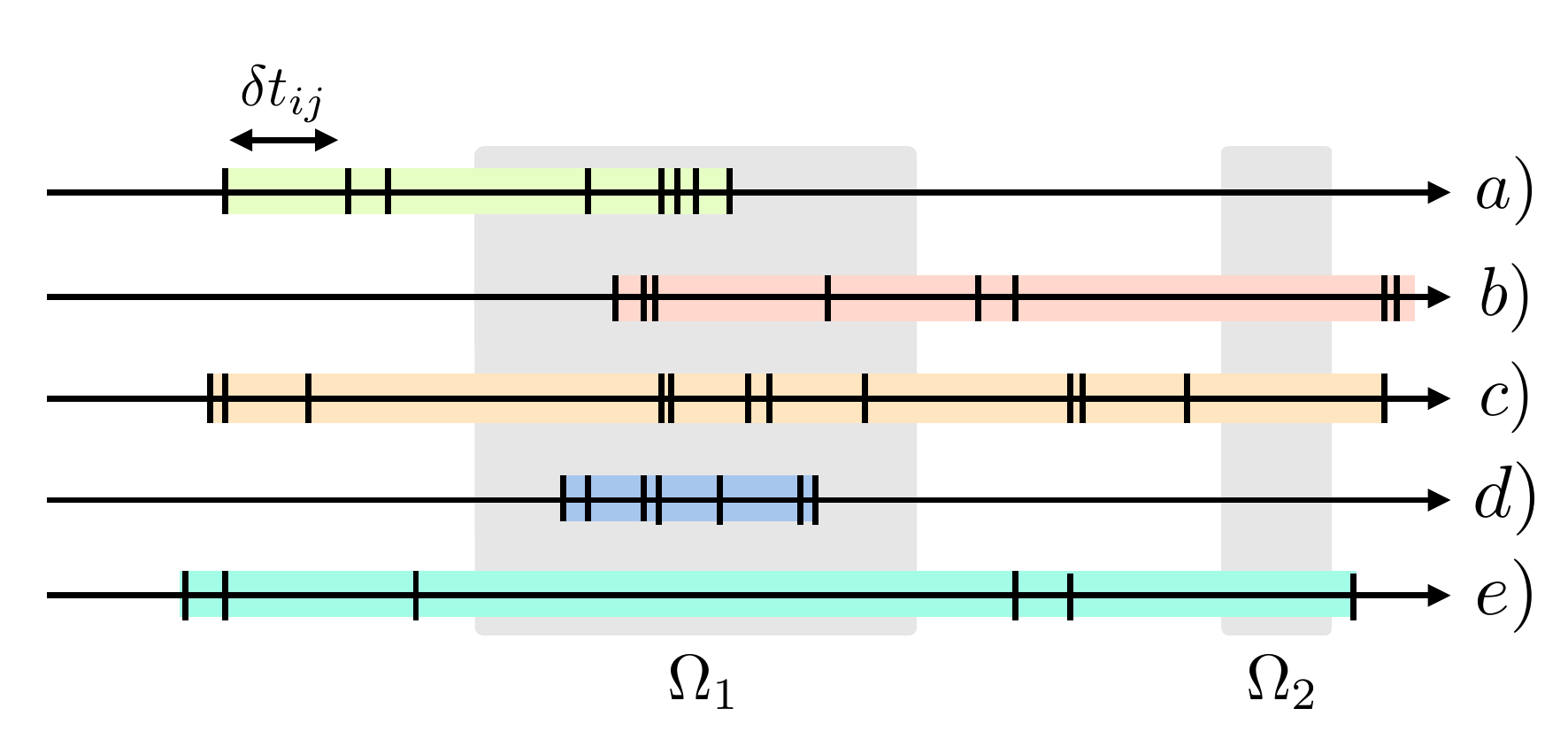}
\caption{Entanglement between bursty dynamics of ties and their formation/decay.  Timelines of calls within 5 given ties. $i\leftrightarrow j$. The gray areas show two different observation periods. If the observation period is very small compared with typical inter-event time $\delta t_{ij} $ (e.g. $\Omega_2$) we might underestimate the ties present. For longer observation periods $\Omega_1$ we need large longitudinal databases to assess that links are created (e.g. link b) or destroyed (e.g. link a). }\label{fig:tiedynamics}
\end{center}
\end{figure}

Mechanisms such as those discussed above should leave traces in empirical data on the dynamics of ties, in the shape of deviations from randomness or invariant features, and many studies in the recent years have attempted to uncover the salient statistical properties of tie dynamics. This has only been possible recently, because the typical time scale of tie dynamics (months) requires longitudinal databases with long periods of observation (years), see Figure \ref{fig:tiedynamics}. On top of that, while there is an explicit friending procedure on some communications platforms (e.g. Facebook), in most situations such as with CDRs tie formation or decay has to be estimated from the initiation and termination of activity within the tie. As we have shown before, communication events are bursty, and a long inactivity period can be mistaken as an absence of the tie. This could be alleviated by using longer and different observation windows (typically around 6 months) \cite{miritello2013limited,holme2003network}, since short time windows can underestimate the main structural properties of the network while overestimating tie dynamics due to the bursty activity within links \cite{krings2012effects,hidalgo2008dynamics}.

\subsection{Dynamics of tie creation and destruction}

In Ref.~\cite{miritello2013limited}, Miritello \emph{et al.}~studied the dynamics of formation and decay of individual links, using a large longitudinal database (19 months) of mobile phone records. They found that as other human activities, link formation/decay events also happen in bursts, e.g. rapid successive creation/destruction events of ties are separated by longer periods of inactivity. Despite this bursty behavior, a strong cutoff in the distribution of inter-event times was found, suggesting that there is a typical time scale of tie dynamics: in Ref. \cite{miritello2013limited} it was observed that on average around one tie is created/destroyed per month in human communication. The existence of this time scale even at the individual level implies that social neighbourhoods change linearly in time. However, as  mentioned before, not all links are equally likely to decay. Miritello \emph{et al.} \cite{miritello2013limited} found that after 6 months of activity the {\em persistence} of social neighborhoods (i.e.~the fraction of an individual's links that remain active during those 6 months) was around 75\%, compared to the expected 50\% of a null model in which each individual's link is equally probable to disappear. Similar long-term persistence of some links was found in Burt's study of 4 years of relationships of individuals in a financial organization \cite{burt2000decay}.

Information diffusion is also affected by the dynamics of tie creation/destruction. Several works have found that in general tie dynamics alone (without considering the bursty nature of events within each tie) slows down the propagation of information \cite{miritello2013temporal, holme2014birth} when compared with null models in which links are usually taken as static. This is due to the unreal assumption that all links observed in a period of time are capable of transmitting information throughout that observation time. The high link turnover observed in human communication suggests that the static picture of human ties overestimates the connectivity potential of individuals.

\subsection{Social strategies and persistent patterns}

\begin{figure*}
\begin{center}
\includegraphics[width=0.99\linewidth]{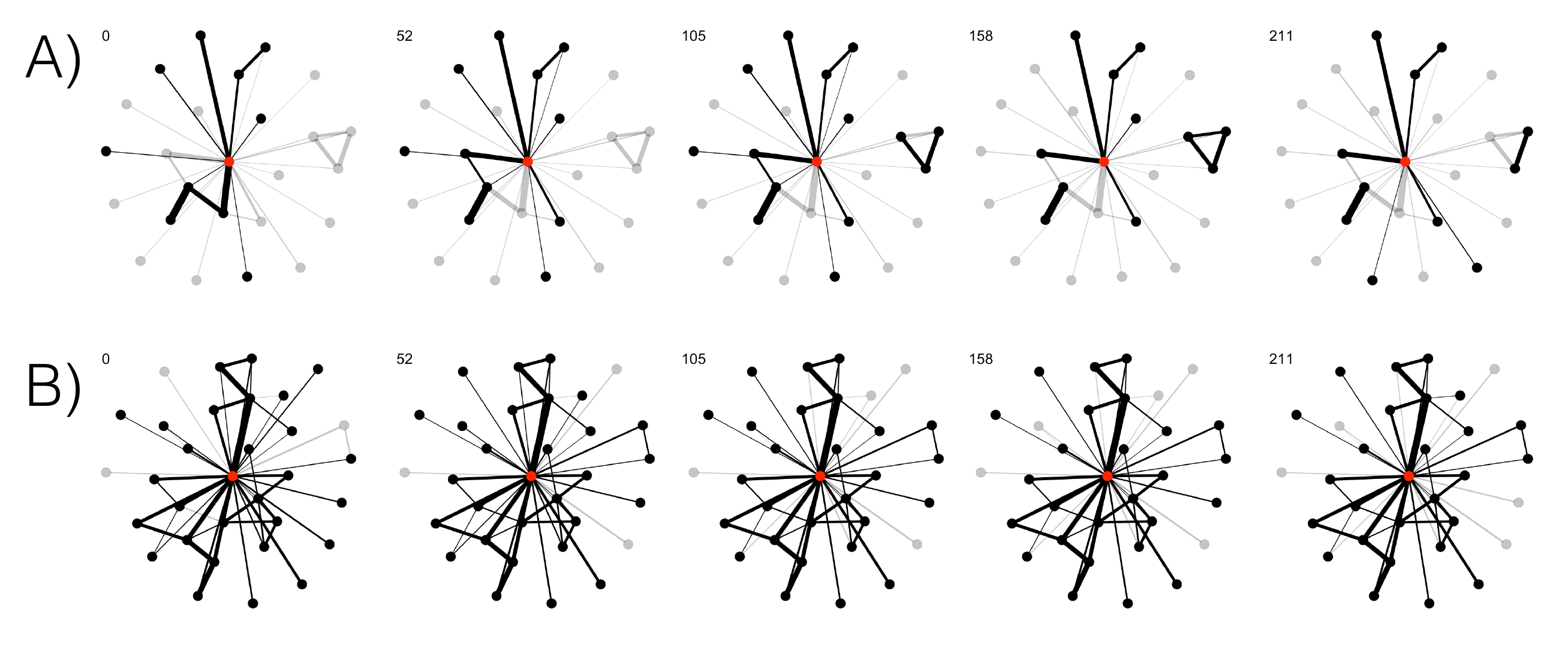}
\caption{Snapshots of the active links (black) in the neighborhood of two different individuals (red symbols) at 5 equally-spaced times during 6 months. Inactive links at a given instant are in gray.  A) user behaves according to the {\em explorer} strategy, while B) follows the {\em keeper} strategy. Link width is proportional to the number of calls. Note that both users present a similar frequency-rank relationship (signature) along the 6 months}\label{fig:egonetworks}
\end{center}
\end{figure*}

Even though ties are continually created and destroyed by individuals at a fast (monthly) pace (their {\em social activity}), it was found in \cite{miritello2013limited} than the rates of creation and destruction are similar for each individual. This implies that the number of active ties at a given instant (the {\em social capacity} of an individual) is almost constant in time. Different combinations of capacity and activity were found that define for each individual a dynamical strategy of communication: while {\em social explorers} have large levels of activity compared to their communication capacity resulting in a fast turnover of their neighborhood, {\em social keepers} activate/deactivate a smaller number of connections compared to their capacity and their social neighborhood is mostly stable (see Fig.~\ref{fig:egonetworks}). Those dynamical communication strategies depend on the age and gender of individuals, with both capacity and activity decreasing as a function of age and being larger for men than women. Finally it was found that there was a significant assortativity of social strategies, meaning that social explorers/keepers tend to gather. These findings render a dynamical picture of the network with very different rhythms of evolution: highly static areas of social keepers live together with extremely volatile groups of social explorers.

Social strategies  also have an impact on an individual's capacity to access information that is being propagated in a network. Using similar SI simulations as mentioned previously, Miritello {\em et al.} \cite{miritello2013limited} found that (for a fixed number of different contacts), social keepers received (i.e., became infected with) the information faster than social explorers. This result suggests that the information access benefits of diverse ties of social explorers are outweighted by their short time lifespan, resulting in a net delay in access to information from individuals activating them.

In Ref.~\cite{Saramaki2014}, Saram\"aki \emph{et al.} have pointed out another feature of individuals' networks that remains invariant even though there is network turnover -- the frequency-rank relationship of numbers of calls to others, called their \emph{social signature}. This means that for a given individual, the fraction of calls targeted to each acquaintance depends on how highly they rank in that individual's network in terms of call numbers, not their identity. Hence, each individual has a characteristic social signature -- e.g., placing a higher fraction of calls to the top 2 acquaintances, or sharing calls more evenly among everyone. These results were obtained using 18 months of data for 24 students who finished high school and went to university or work, which guaranteed rather high levels of turnover in their social networks. 

Ref.~\cite{perra2014} looked at egocentric network evolution from a different point of view -- that of new communication events being associated with existing or newly appearing ties. They found a universal formula for the probability of observing new versus existing ties.


\section{Community and network evolution}

At even longer time scales -- years -- we find that some parts of the network or even the network itself change dramatically. Palla {\em et al.}\ \cite{Palla2007} found with CDR and other data that social groups or communities within networks have their own dynamics. Concurrent tie formation/decay events inside and around those communities give rise to growth, contraction, merging, splitting, and birth or death of communities. Interestingly, Palla {\em et al.} also found that while larger communities are on average older, they also have higher rates of change. Thus, large communities survive because of a continuous turnover of their members. For example, there is a sustained flow of individuals joining communities and leaving them. Backstrom {\em et al.} found with data on online and co-authorship networks that membership is contagious: the probability to join a community depends on the number of friends previously in the community \cite{backstrom2006group}. The typical scale of community dynamics depends on their size, although there are merging or splitting events in communities that happen in short periods of time (two weeks in \cite{Palla2007}), suggesting that community evolution resembles the punctuated equilibrium of biological species. 

In order to discuss communication network dynamics on even larger scales, we have to take a small detour from the focal topic of call networks and enter the online world. The recent explosion of online social networks has not only allowed to study how successful, massive networks grow in time \cite{leskovec2008microscopic}, but also to perform an autopsy on the late stages of unsuccessful ones \cite{garcia2013social}. By studying the growth of networks like LinkedIn, Decilious or Flickr, Leskovec {\em et al.}~\cite{leskovec2008microscopic} found that nodes and edges do not arrive linearly in time. Rather, their growth dynamics accelerates after the first year, probably due to non-linear social effects in the adoption contagion or external mass media influence as Toole {\em et al.}~found for Twitter \cite{toole2012modeling}. Despite this non-linear effect, tie dynamics happen mostly following the well-known processes of preferential attachment, triadic closure \cite{leskovec2008microscopic} or geographical homophily \cite{toole2012modeling}.

In addition to network growth, there is also network decline, or at least membership turnover. When it comes to social networks built from CDR data provided by a single carrier, network turnover is mostly related to  some users eventually deciding to abandon the carrier and switch to another carrier. This is known as \emph{churn}. The rate at which client churn happens is large (at around 2\% per month for wireless carriers in the US). This large turnover of the networks can be analyzed using CDRs: for example Dasgupta {\em et al.} found that as with community evolution, the decision to leave the carrier is highly correlated to the number of friends that have previously left the carrier \cite{dasgupta2008social}. A similar form of social contagion is behind adoption of products or services in networks: using 2 years of CDR data, Sunds{\o}y {\em et al.}~ \cite{sundsoy2010product} found that product adoption spreads through the social network of clients. 

The decline of entire networks has been studied with data on some online social networks that shrunk after their growth phase. Garcia {\em et al.} \cite{garcia2013social} studied two years of decline of networks like Friendster to investigate the causes and mechanisms behind their failure. The found that as with mobile call networks, individual decisions regarding participating in or leaving a community or network are, to a large extent, determined by the number of one's friends in the social network and their own engagement with the community/network. Thus, a fraction of one's friends leaving the network/community can trigger one to leave, resulting in further cascades of leaving events and eventually in the network shrinking and finally ceasing to be. This cascading process accelerates in time and thus network decay can eventually be really fast. For example, the Friendster network shrunk from 60 million users to 10 million users in a year probably because of a cascading process (but also because of mass media and competition with other online social networks). 

Finally, it is of interest to note that although mobile call networks are shaped by a number of processes from dynamics of egocentric networks and communities to customer churn, key network-level characteristics such as connectivity and tie strength distributions appear to remain stationary over long periods of time~\cite{krings2012effects}. That is, the details of the networks change, but the big picture does not.

\section{Future outlook}
As seen above, mobile telephone call records have allowed us to better understand human communication dynamics, and through that, the dynamics of social networks. Where, then, is this field heading? Interest in mobile telephone data still certainly growing, as seen in the success of e.g.~the Netmob conference on mobile phone data set analysis (\emph{www.netmob.org}), and there are plenty of open issues to be addressed. At the same time, there is an ever-expanding diversity of communication channels which necessitates approaches that do not rely on a single source of data. It may even be that the golden age of CDR-based research is slowly coming to its end, as  the younger generations adopt new channels of communication even for voice (Skype, voice over IP). Mobile communication via such channels is only seen as
data traffic and details such as recipients of messages are not recorded by the mobile telecom. With a multitude of channels operated by different companies, data collection on massive scales becomes difficult or impossible. This may necessitate collecting research data via smartphone apps from consenting volunteers; although the numbers of participants will necessarily be smaller, this may be compensated by an increase in data quality and coverage of multiple channels.

In the following, we will attempt to identify some emerging themes and trends.

\subsection{Experiments: From big data to deep data}

The advantage of using CDRs, extracted from mobile operators' billing systems, is the sheer size of data, both in terms of numbers of users and in terms of call events. However, at the same time, such data is necessarily shallow~\cite{Lehmann2014}. Because of privacy reasons, information on phone users is very limited (\emph{e.g.}~age, gender) or not available at all. Furthermore, mobile telephone calls represent only one channel in an ever-expanding multitude of electronic communication channels. Yet data originating in mobile telephone operator billing system does not contain information on any other channels, with the exception of text messages (whose use has already declined in the younger generations). 

There is only one practical way of sorting out the above difficulties, and that is collecting research data already at the user end, for example with a smartphone app designed for the purpose. As mentioned above, this may also be the only viable option in the long term because of the increasing diversity of communication channels results in a lack of coverage by CDRs. Then, instead of using anonymised large-scale data, one needs to go for volunteer users. In other words, the data collection phase becomes its own project -- an experiment designed by researchers. This necessarily limits the number of studied individuals, as it is hard to scale up any experiment to the level of entire nations that CDRs cover. Participant retention is another problem, especially for longitudinal experiments with intended time spans of years instead of months. In the earliest experiments that combine call records with other types of data (such as GPS positioning and Bluetooth proximity)~\cite{EagleReality2006,Kiukkonen_2010,Aharony2011,Karikoski2011}, the numbers of participants have been of the order of $\sim 100$. An effort that is larger by one decade -- the Copenhagen Networks Study with $N\sim 1000$ -- is currently approaching the end of its data collection phase~\cite{Lehmann2014}.

App-based collection of data allows recording a number of data streams from GPS positioning to usage of other apps and communication channels. On top of that, it is possible to apply the traditional method of social sciences: actually \emph{ask} the users what they are doing, why, and how do they feel about it, via pop-up surveys. These, together with psychological profiling (\emph{e.g.}~standard questionnaires for personality traits, see ~\cite{Lehmann2014}) provides a far more detailed picture on the users than what can be obtained from mobile telephone operators. Additionally, surveys can provide valuable information on the nature and closeness of the social ties captured by electronic communication, since different types of ties play different roles in network structure (see, \emph{e.g.},
~\cite{Backstrom2014}). Given data sets with temporal information, ground truth, and enough statistical power, it might even be possible to associate features of call time series with the nature of social ties, something that could then be applied to unlock features of larger data sets.

\subsection{From aggregates to individuals}

In the early days of social network analysis with large databases, focusing on structural properties of networks and disregarding details such as possible differences between individuals was the norm (except perhaps for issues such as broad connectivity distributions of individuals). Likewise, the social network modelling paradigm has mainly been that of social atoms: in a typical agent-based model, each node follows exactly the same rules as everyone else, and the aim is to see whether such minimalistic assumptions can already explain empirically observed features. However, it is evident that much important information is lost when disregarding individual differences and relying on system-level summary statistics on network structure and dynamics. In the worst case, this can lead to so-called \emph{ecological fallacy}~\cite{Robinson1950}, where statistical dependencies seen at the system level are falsely attributed to the individuals comprising the system too. The opposite is also possible: much interesting and important variation may be hidden behind flat system-level averages. It should be noted that statistical physicists are especially vulnerable to this problem, being used to system-level statistics describing the behaviour of large numbers of identical elements. 

Moving beyond the aggregate level requires more detailed information than knowledge of network structure alone (which is probably why so many studies have remained there). In addition to harder-to-collect experimental data (see above), when it comes to CDRs, such information can be available in the time domain. Even when the source data contains no direct information on each individual's attributes, their behavioural patterns manifested in temporal event sequences and link dynamics may allow distinguishing between different types of individuals, or inferring some personal traits and features. The discussion on social strategies (keeper, explorer) above is a good example of this. Known attributes of individuals such as gender and age have also been seen to affect their temporal communication patterns in Ref.~\cite{Kovanen2013}. When call data collected in experiments is augmented with extra information such as surveys and psychological profiling, an entirely new set of possibilities opens up. As an example, correlations beetween extraversion and number of calls contacts have been found in the analysis of CDRs ~\cite{staiano2012friends,de2013predicting,Lehmann2014}, similarly to Facebook friends~\cite{Quercia2012}. 

In addition to data analysis, we expect that the next generation of improved social network models will build on diversity in individual behaviour instead of identical agents.

\subsection{Data sharing}

Perhaps the biggest problematic issue related to mobile call datasets is that it is typically impossible to publicly share data. Most results on mobile call networks are outputs of a similar pipeline: a mobile telecom operator agrees to provide anonymised data to a group of researchers under strict non-disclosure agreements and under the condition that any results to be published must be scrutinised by the company first to avoid publishing commercial secrets. Anonymisation is typically achieved simply by replacing all phone numbers with surrogate keys. Since the data cannot be shared, one of the main principles of the scientific method, reproducibility of results, is violated: the only way of checking someone's results is to succeed in obtaining similar data from a telecom, and even then the results may differ because of \emph{e.g.}~sampling or cultural issues. 

The problem with releasing detailed call data with time stamps is that it is very difficult to guarantee that users cannot be re-identified even in anonymized data. As an example, it is easy to identify oneself by matching the time stamps found in the call log of one's own mobile phone with time stamps in CDR data; this also reveals everyone who has been called. In fact,  structural network  information without time stamps may already be enough, see \cite{Backstrom2007}. Inclusion of tower location data brings even more problems~\cite{Song2010}, since individuals can be identified from their frequented locations~\cite{DeMontjoye2013}. However, if the data has been collected experimentally, with consenting volunteers, it may be possible to release at least parts  of it, depending on a number of issues such as agreements with participants and the level of anonymity in the released data. 

Because of the above problems, there are very few publicly shared data sets -- the MIT reality mining data includes call logs~\cite{EagleReality2006}, Wu \emph{et al.}~\cite{Wu2010} comes with three sets of data on anonymised, time-stamped text messages, and Saram\"aki \emph{et al.}~\cite{Saramaki2014} provides three sets of egocentric networks aggregated over 6 months each. A notable exception in data sharing is the different data challenges that have taken place in the last years, such as the Nokia Mobile Data Challenge by Nokia \cite{nokia}, the two D4D Challenges by Orange~\cite{D4D} or the Telecom Italia Big Data Challenge \cite{telecomitalia}. In these challenges, anonymised samples of call and mobility data are made available for researchers for limited time upon request. The aim is to use the data for research projects that e.g.~have a development dimension like in the D4D challenges (Ivory Coast in 2013, Senegal in 2014) or address applicability in sectors like energy, weather, public and private transport, and social network studies. Here, privacy issues have been addressed by a number of techniques: small samples, aggregation and coarse-graining, and added noise. It is worth noting that although the data is available to any researcher who wishes to participate in the competition, it still comes with a non-disclosure agreement and its use is limited to the competition. A remarkable exception is the data by Telecom Italia, who have opened their challenge data for  reuse \cite{telecomitaliaopen}. 

Would it then be possible  at all  to share mobile call data, without aggregating out too many details and while still preserving privacy? There are no readily available solutions to this problem. The concept of \emph{homomorphic} encryption (see, \emph{e.g.},~\cite{Gentry2009}) has been discussed in  contexts such as cloud security. In this scheme, a limited set of analysis operations can be conducted on data that is already encrypted. However, the viability of such a scheme for CDR analysis is uncertain.
Another possibility might be not to release the data itself, but let researchers access to it through an Application Programming Interface (API) that allows using highest-resolution data in computations, but only provides aggregated results, along the lines of \emph{openPDS}~\cite{montjoye2014}.

\section{Acknowledgements}
Support by the Academy of Finland, project no: 260427 and Spanish MINECO Grant FIS2013-47532-C3-3-P are acknowledged. We thank Manuel Cebri\'an, M\'arton Karsai, Sune Lehmann and Giovanna Miritello for useful discussions.

\bibliography{minireview} 

\end{document}